\documentclass{elsart}

\usepackage{graphicx}
\usepackage{epsfig}

\usepackage{amssymb}
\usepackage{color}

\begin{document}

\begin{frontmatter}



\title{A Model of Nuclear Recoil Scintillation Efficiency in Noble Liquids}
\author[usd]{D.-M. ~Mei\corauthref{cor}}
\corauth[cor]{Corresponding author.}
\ead{Dongming.Mei@usd.edu}
\author[usd,ccnu]{Z.-B. ~Yin\thanksref{now}},
\thanks[now]{Permanent Address: Institute of Particle Physics, 
Huazhong Normal University, Wuhan 430079, China}
\author[lanl]{ L.C.~Stonehill},
\author[lanl]{A.~Hime}

\address[usd]{Department of Physics, The University of South Dakota, 
Vermilion, South Dakota 57069}
\address[ccnu]{Institute of Particle Physics, Huazhong Normal University, 
Wuhan 430079, China}
\address[lanl]{Los Alamos National Laboratory, Los Alamos, NM 87545}
\begin{abstract}
Scintillation efficiency of low-energy nuclear recoils in noble liquids 
plays a crucial role in interpreting results from some direct searches for Weakly Interacting 
Massive Particle (WIMP) dark 
matter. However, the cause of a reduced scintillation efficiency relative to electronic recoils 
in noble liquids remains unclear at the moment. 
We attribute such a reduction of scintillation efficiency to 
two major mechanisms: 1) energy loss 
and 2) scintillation quenching. The former is commonly
described by Lindhard's theory and the latter 
by Birk's saturation law. We propose to combine 
these two to explain the observed reduction of scintillation yield for nuclear recoils in noble liquids. 
Birk's constants $kB$ for argon, neon and
xenon determined from experimental data are used 
to predict noble liquid scintillator's response to low-energy nuclear recoils 
and low-energy electrons. We find that energy loss due to nuclear 
stopping power that contributes little to ionization and excitation 
is the dominant reduction mechanism in scintillation efficiency for nuclear recoils, but that significant additional quenching
results from the nonlinear response of scintillation to the 
ionization density.
  
\end{abstract}
\end{frontmatter}

\maketitle

\section{Introduction}
Noble liquid scintillators such as liquid xenon~\cite{xenon10}, 
argon~\cite{bh,warp}, and neon~\cite{dan} are expected to be 
excellent targets and detectors for direct dark matter detection experiments 
searching for Weakly Interacting Massive Particles (WIMPs), 
which may constitute the dark matter in the 
universe~\cite{dns, wfr, mwg, gjm}. 
These experiments measure scintillation light 
induced by low-energy nuclear recoils due to elastic
scattering of WIMPs. Absorption of nuclear recoil energy in noble liquid scintillators
produces excitons and electron-ion pairs along the track. 
 Free excitons collide
with ground states to form excited molecules (excimers) through 
\begin{equation}
\label{eq:mod1}
X^{*} + X \rightarrow X_{2}^{*},
\end{equation}
 where X stands for
any type of noble liquid.
Free ions undergo collision, recombination and deexcitation processes, 
\begin{equation}
\label{eq:mod2}
 \left\{ \begin{array}{ll} 
X^{+} + X \rightarrow X_{2}^{+},\\
X_{2}^{+} + e^{-} \rightarrow X^{**} + X,\\
X^{**} \rightarrow X^{*} + heat, \\
X^{*} + X \rightarrow X_{2}^{*}, \\
\end{array}\right.
\end{equation}
to form excimers. The excimers then decay radiatively from the lowest-excited molecular states $^{1}\Sigma_{u}^{+}$ and 
 $^{3}\Sigma_{u}^{+}$ to the repulsive ground state $^{1}\Sigma_{g}^{+}$.

It is well known that noble liquid scintillators 
have reduced scintillation yield for low-energy nuclear recoils compared to electronic recoils~\cite{xe1,xe2,xe3,ar,ar1,ne}. Only
a fraction of the energy loss results in ionization and atomic 
excitation. Moreover,
 high ionization density undermines recombination 
of electron-ion pairs and reduces scintillation light yield. 
The relative scintillation yield, defined as the ratio of the 
numbers of photons emitted from pure noble liquids in 
nuclear and electronic recoil events at the same energy, 
is a good measurement of the nuclear recoil scintillation efficiency, 
$q_{f}$, determined by the visible deposited energy over 
the true recoil energy. In the case of 
electrons and $\gamma$-rays,  almost all the energy loss by ionization 
is converted into scintillation light through electron-ion recombination,
so the relative scintillation efficiency is assumed approximately equal to 1
in the absence of an electric field. 
But in the case of nuclear recoils, $q_{f}$ is much smaller than 1 
and can vary as a function of nuclear recoil energy. 

The scintillation efficiency plays an important role in 
the direct detection of WIMPs. In the design of a new experiment, 
the scintillation efficiency is related to the detection threshold and hence to 
the background level and ultimate sensitivity. In the interpretation of an experimental 
result, the scintillation efficiency is crucial to the determination of WIMP 
mass and WIMP-nucleon cross section. 
The nuclear recoil scintillation efficiencies for liquid xenon, argon, and neon 
have been measured~\cite{xe1, xe2, xe3, ar, ar1, ne} using neutron sources. 
The detector is usually calibrated  
with well-known $\gamma$-rays, 
such as 122 keV and 133 keV lines from a $^{57}$Co source. 
The relative scintillation efficiency for $\gamma$-rays, $\epsilon_{\gamma}$, is defined as the 
visible energy divided by the incident $\gamma$-ray energy, and  
is assumed to be 1. As discussed later in this paper, 
this assumption is valid for electronic recoil energy 
above 20 keV in the noble liquids under investigation. 
The nuclear recoil scintillation efficiency is determined by the ratio of the nuclear recoil
visible energy using the electron-equivalent energy calibration to the true recoil energy $E_R$, 
$q_{f} = E_{R}^{\rm{vis}}/E_{R}$. 

The measurements are usually compared to either Lindhard's theory~\cite{lind} 
or Hitachi's treatment~\cite{hit}. It was found that Lindhard's theory 
alone can not well explain the observed behavior in the data. 
The alternative explanation, Hitachi's treatment, states
that a biexcitonic quenching mechanism can occur before the free excitons
self-trap when the excitation density is very high. This explanation  
can agree reasonably well with the xenon data~\cite{xe1}, but has not been applied to 
explain low-energy recoil data for neon or argon. Moreover, there are other possible quenching processes that could contribute
to the reduction of scintillation efficiency for low-energy nuclear recoils. These include collisions (via the Penning process~\cite{cla}) 
between two excited molecular states (excimers) to form one excited state 
and one ground state~\cite{cap},  and supereleastic 
collisions that quench the singlet states to triplet states~\cite{dcl}.

A more universal description of the reduced scintillation efficiency for nuclear recoils
is preferred for all noble liquid scintillators. Since the proposed quenching mechanisms are all dependent on the density 
of the ionization and excitation track left by the recoiling nucleus, it is possible to form a combined model of these 
mechanisms without incorporating details of the relative contributions of the different mechanisms.  Birk's saturation law for 
organic scintillators~\cite{birk} provides a convenient description of the dependence of scintillation quenching on ionization density.
 In this study, we apply Birk's law to noble liquids, offering a conventional way to determine the total scintillation efficiency of
nuclear recoils by measuring Birk's constant ($kB$).  It has been shown that the luminescence intensity in the noble gas scintillator 
depends solely on the energy density and is independent of the kind of the particle~\cite{att, attt}. This is to say that a measurement of $kB$ for noble 
liquids will allow understanding of the relative scintillation efficiency for nuclear recoils induced by neutrons, alphas, and other heavy isotopes.  
This is a very valuable way to determine the scintillation efficiency for nuclear recoils induced by all types of particles in noble liquid scintillators.  
Furthermore, this method allows determination of the relative scintillation efficiency for nuclear recoils in noble liquids by measuring the constant $kB$ 
with $\gamma$-rays.  This is a much easier measurement compared to the nuclear recoil measurements with neutron and alpha sources.

In this paper, we propose a model to combine Lindhard's theory and 
Birk's saturation law to describe the reduction in scintillation efficiency observed in 
noble liquid scintillators.  We describe these two reduction mechanisms in Section~\ref{sec:quen1} 
and~\ref{sec:quen2}, respectively. The model combining these two reduction mechanisms 
is presented in Section~\ref{sec:com} and its predictions  
are compared to experimental data 
in Section~\ref{sec:comparison}. The scintillation efficiency ($\epsilon_{\gamma}$ 
as a function of recoil energy) for very low-energy electrons and $\gamma$-rays is discussed briefly in Section~\ref{sec:ele}. 
Finally, we summarize our conclusions in Section~\ref{sec:con}. 

\section{Reduced Ionization Energy by Nuclear Collisions}
\label{sec:quen1}

When a neutron or WIMP scatters elastically off a noble liquid atomic 
nucleus, 
the recoiling nucleus then loses its energy by colliding with electrons 
and nuclei within the detector.
This nuclear recoil process involves the competition between, on 
the one hand, energy transfer to atomic electrons and, 
on the other hand, energy transfer to translational motion
of atoms. The total rate at which the recoiling nucleus loses energy 
with respect to distance ($dE/dx$) is dependent
on the medium through which it travels, and is also called the stopping power. 
At low energies, the total stopping power of the noble liquid atom 
consists of electronic and nuclear 
stopping power. The electronic stopping power is the amount of energy 
per unit distance that
the recoiling nucleus loses due to electronic excitation and ionization 
of the surrounding noble liquid atoms.
The nuclear stopping power is the energy loss per unit length 
due to atomic collisions that contribute to the kinetic 
energy (thermal motion) of the noble liquid atoms, but that do not result 
in internal excitation of atoms. The proportion of electronic 
to nuclear stopping power depends on the recoil energy of the nucleus. 
If the recoil energy were very large, the nuclear stopping power 
would be very small compared to the electronic stopping power. 
However, in the energy range of WIMP-nucleus elastic 
scatterings, the nuclear stopping power plays a 
significant role in the energy loss of the recoiling noble liquid nucleus. 
J. Lindhard {\it et al.}~\cite{lind} discussed in detail 
the theory of energy loss of low-energy nuclei.

\begin{figure}[htb!!!]
\includegraphics[angle=0,width=12.cm] {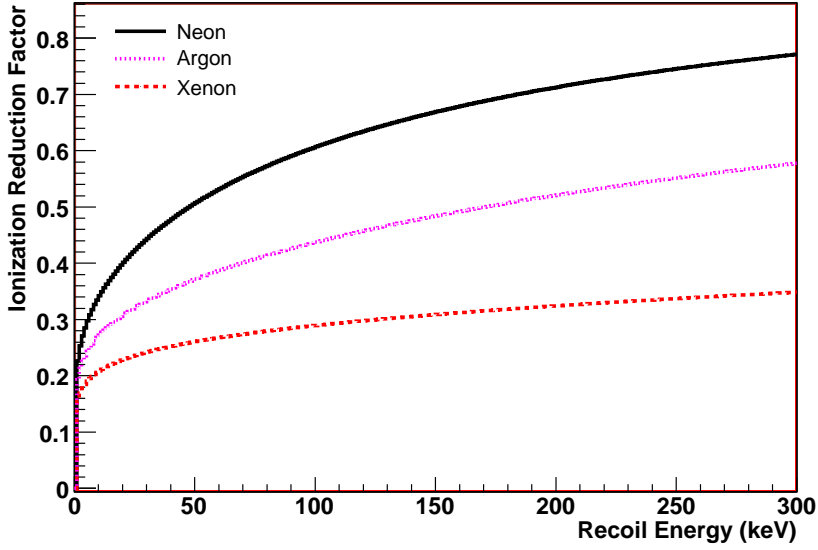}
\caption{\small{
The ionization energy reduction factor according to Lindhard's theory versus 
the recoil energy in liquid neon (black line), argon (red dotted line) and xenon (magenta dashed line).}}
\label{fig:lindfirst}
\end{figure}

Supposing that the recoiling nucleus loses all of its energy in the detector, 
the total energy loss can be expressed in terms of the losses
due to the electronic stopping power $\eta$ and nuclear stopping 
power $\nu$ as~\cite{lind}
\begin{equation}
\label{eq:loss}
E_{R} = \eta(E_{R}) + \nu(E_{R}),
\end{equation} 
where $\eta$ and $\nu$ are both functions of recoil energy $E_{R}$. 
As only the portion of the energy lost in 
electronic excitation or ionization will result in the creation of excitons and electron-ion pairs  
in the noble liquids, the fraction defines
an ionization energy reduction factor ($f_{n}$) due to losses to the nuclear stopping power
\begin{equation}
\label{eq:frac}
f_{n}(E_R) \equiv \frac{\eta(E_{R})}{E_{R}} 
= \frac{\eta(E_{R})}{\eta(E_{R}) + \nu(E_{R})}.
\end{equation}
As the total stopping power is
\begin{equation}
\label{eq:stop}
(\frac{dE}{dx})_{\rm{tot}} = (\frac{dE}{dx})_{\rm{elec}} 
+ (\frac{dE}{dx})_{\rm{nucl}},
\end{equation}
$f_n(E_R)$ can then be determined by the ratio of two integrals
\begin{equation}
\label{eq:inte}
f_n(E_{R}) = \frac{\int_{0}^{E_{R}}(dE/dx)_{\rm{elec}} dE}
{\int_{0}^{E_{R}} ((dE/dx)_{\rm{elec}} + (dE/dx)_{\rm{nucl}}) dE} .
\end{equation}
To present $f_n$ as a function of recoil energy, 
the integrals above should be evaluated 
for each possible recoil energy. 
Lindhard {\it et al.}~\cite{lind} represents $f_{n}$ as
\begin{equation}
\label{eq:quen}
f_{n} = \frac{kg(\varepsilon)}{1+kg(\varepsilon)},
\end{equation}
where, for a nucleus of atomic number $Z$, 
$\varepsilon = 11.5E_{R}$ (keV) $Z^{-7/3}$, $k = 0.133Z^{2/3}A^{-1/2}$,
and $g(\varepsilon)$ is well fitted by: 
$g(\varepsilon) = 3\varepsilon^{0.15}+ 0.7\varepsilon^{0.6} + \varepsilon$.
Fig.~\ref{fig:lindfirst} shows this ionization energy reduction factor for noble liquids 
from Lindhard's theory.

\section{Reduced Scintillation Yield due to High Ionization Density }
\label{sec:quen2}
\subsection{Birk's Saturation Law }
\label{sec:app}
The passage of a particle in a noble liquid produces a structured track along its path that is conveniently described in terms of a core 
and a penumbra~\cite{moz}.
The penumbra that surrounds the core is a low ionization density zone.  The core is expected to be a high ionization density zone, 
so ionization density dependent quenching caused by biexcitonic collisions or the Penning process is likely to occur there.
On the one hand, the free excitons can be self-trapped to form 
excimers, for example, Ar$_{2}^{*}$, that then fluoresce to the lowest states. 
On the other hand, 
the free excitons can diffuse and undergo biexcitonic quenching, or the excimers can further collide with each other via the Penning process,  
with a probability of occurrence that depends on 
the density of free excitons produced, which is proportional to ionization density. Therefore, 
the biexcitonic collisions and the Penning process act as quenching 
agents for the excitons produced by the 
ionization along the track. It is well known that 
the number of excitons and electron-ion pairs produced per unit path length is proportional to the
electronic energy loss $dE/dx$~\cite{kub}, with a proportionality constant that will be designated as $A$.   
The local concentration of the core is also proportional 
to the ionization density, and is given by $B$$dE/dx$.
The overall collision probability in the core is given by $k$. Thus, the specific fluorescence can be expressed by 
\begin{equation}
\frac{dS}{dx} = \frac{A \frac{dE}{dx}}{1 + kB \frac{dE}{dx}}.
\label{eq:birk0}
\end{equation} 
Eq.~\ref{eq:birk0} is called Birk's saturation law~\cite{birk}, which describes 
the relative scintillation response of scintillators to an ionizing particle of any energy. 
The values of $A$ and $kB$ can be determined experimentally.

According to Eq.~\ref{eq:birk0}, the scintillation photon 
yield is reduced at high ionization density. Thus, we define a 
quenching factor 
\begin{equation}
\label{eq:birk2}
f_{l} = \frac{1}{1 +  kB\frac{dE}{dx}},
\end{equation}
which is related to the electronic stopping power
$dE/dx$ induced by nuclear recoils. 
 
\subsection{Electronic Stopping Power for Heavy Ions}
\label{sec:signal}

For ions heavier than protons, the electronic stopping power in a given 
material can be calculated from the stopping power in the same material 
based on the ``heavy ion scaling rule'',  
\begin{equation}
S_{A} = (\zeta Z_1)^2 S_{p} ,
\label{eq:S_A}
\end{equation}
 
\noindent
where the stopping power $S_{p}$ for protons is determined at the same 
velocity as the stopping power $S_{A}$ for heavy ions. $\zeta Z_{1}$ is the `effective charge' for ions of atomic number $Z_{1}$. Effective 
ion charges will result from the stripping of bound electrons 
from the ion when moving through a medium. 
Thus, the effective charge fraction $\zeta$ is 
expected to be related to the relative velocity, $v_r$, between the ion
velocity $v_1$ and the velocity of the valence electrons in the medium.
When the valence electron gas of the medium is characterized by the effective
number of electrons that participate in plasma excitations, the velocity of
the valence electrons can be characterized by the Fermi velocity 
$v_F = (3\pi^2 n)^{1/3}\hbar/m$, where $n$ is the electron density and
$m$ the electron mass. $v_r$ is proposed in~\cite{Man81} to be
\begin{equation}
v_r = \left\{ \begin{array}{ll} 
\frac{3}{4}v_F(1+\frac{2}{3}\frac{v_1^2}{v_F^2}-\frac{1}{15}
\frac{v_1^4}{v_F^4}), & \textrm{if $v_1 \leq v_F$} \\
v_1(1+\frac{1}{5}\frac{v_F^2}{v_1^2}), & \textrm{if $v_1 > v_F$} . 
\end{array}\right.
\end{equation}

A formula for the effective charge fraction $\zeta$ has been deduced by Brand 
and Kitagawa~\cite{Bra82} as a good approximation in the region where stopping 
power is proportional to the velocity $v_1$:
\begin{equation}
\zeta = q + 0.5(1-q)\ln\big[1+(\frac{2\Lambda v_F}{1.919v_0a_0})^2\big] ,
\end{equation}

\noindent
where $v_0$ and $a_0$ are the Bohr velocity and radius, respectively. $q$ 
denotes the degree of ionization, which was calculated by applying 
a velocity-stripping
criterion to the Thomas-Fermi model~\cite{Spr91} of the neutral atom 
and is tabulated as a function of the reduced 
variable $y_r = v_r/(Z_1^{2/3}v_0)$ in~\cite{Bra82}. An alternative 
energy stripping criterion was proposed by Mathar and Posselt 
(MP)~\cite{Mat95} 
to explain the dependence of the ionization fraction on the ion velocity and to yield better results especially for higher 
ion velocities. Using the energy definition of 
the Brandt-Kitagawa (BK) model, $q$ is determined by solving numerically the
following equation
\begin{equation}
6 a(1-q)^{2/3}y^2 =\frac{q(6+q)}{7},
\end{equation} 

\noindent
where $a \equiv 0.24005$ and $y\equiv v_1/(v_0Z^{2/3})$. 
The screening radius $\Lambda$, which is used to 
parameterize the charge density profile of the projectile ion, is determined 
by minimizing the total energy of the bound electrons in BK to be
\begin{equation}
\Lambda = \frac{2a(1-q)^{2/3}a_0}{Z_1^{1/3}[1-(1-q)/7]}.
\end{equation}

\noindent
Ziegler, Biersack and Littmark (ZBL)~\cite{zbl} modified 
the ion size parameter $\Lambda$
to include a tabulated factor individual to $Z_1$. The ionization fraction
$q$ is then parameterized as a universal fitting function of $y_r$ as
\begin{equation}
q = 1-e^{(0.803y_r^{0.3}-1.3167y_r^{0.6}-0.38157y_r-0.008983y_r^2)}.
\end{equation}

\begin{figure}[htb!!!]
\includegraphics[angle=0,width=12.cm] {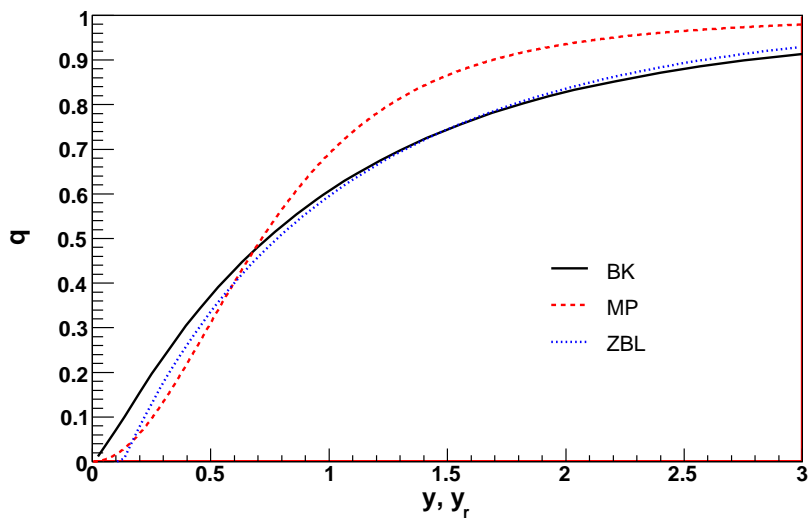}
\caption{\small{
Ionization fraction as a function of variable $y$ or $y_r$ in the BK (black solid line), ZBL (blue dotted line) and MP (red dashed line) models. }}
\label{fig:q_y}
\end{figure}

The ionization fractions from the three different approaches 
described above are reproduced in Fig.~\ref{fig:q_y}
as a function of the variables $y$ or $y_r$. 
In this paper, we adopt the BK formalism to calculate 
the effective charge of ions and the electronic
stopping power values, but the alternative approaches 
are used to estimate the theoretical uncertainties. 

\begin{figure}[htb!!!]
\includegraphics[angle=0,width=12.cm] {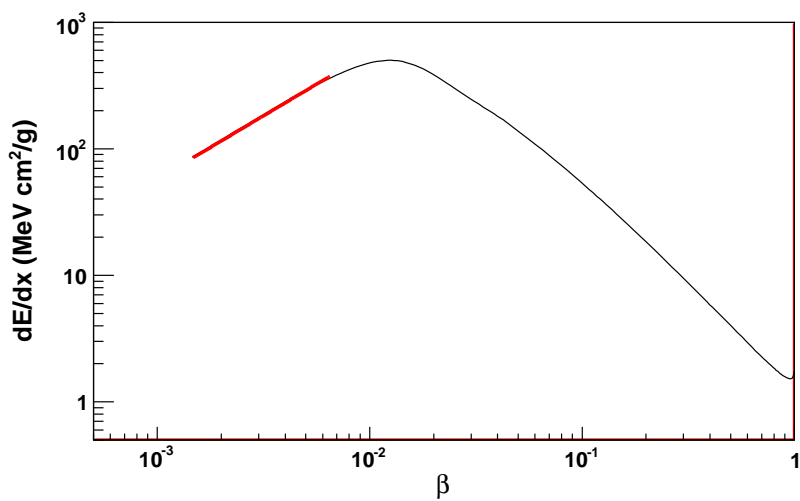}
\caption{\small{
Electronic stopping power for protons in an Ar target versus the 
velocity $\beta \equiv v/c$ of the proton projectiles. 
A proportional function (red thick line) is fitted to 
the low velocity region with a parameter 
of $5.776\times 10^4\pm 349.7$ MeV cm$^2$/g.}}
\label{fig:pStoppingAr}
\end{figure}

We can utilize Eq.\ref{eq:S_A} and the effective ion charge to 
calculate the electronic stopping power, provided that the stopping 
power values for protons in the same material are known. A table of stopping powers and ranges for protons and alpha particles
has been produced by a committee of the International Commission 
on Radiological Units and Measurements (ICRU)~\cite{ICRU49}. 
In the present study, we retrieve the proton stopping powers from 
the PSTAR database~\cite{PSTAR}. 
Fig.~\ref{fig:pStoppingAr} shows the electronic stopping power
for proton projectiles in Ar target as a function of the proton
velocity $\beta \equiv v/c$. According to Lindhard and Scharff~\cite{Lin61},
the stopping power at low energy is proportional 
to the projectile velocity. This is illustrated by 
a proportional function (red thick line) fit in Fig.~\ref{fig:pStoppingAr}.
We will use this fitting function to estimate the stopping power at lower
energy, where it is not covered by the databases, to produce the electronic
stopping power for ions at kinetic energy as low as 1 keV.

\begin{figure}[htb!!!]
\includegraphics[angle=0,width=12.cm] {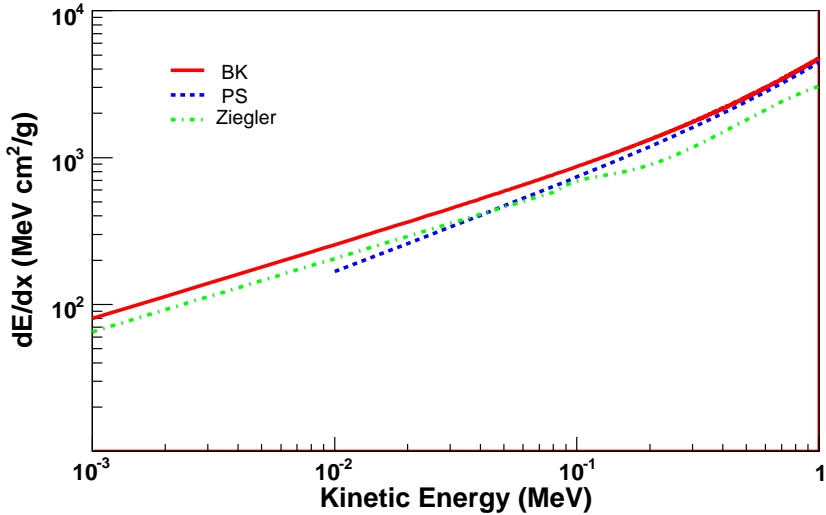}
\caption{\small{
Electronic stopping powers for Ar ions in Ar targets as a function of the
kinetic energy. The red solid line is from the BK model, the blue dashed line from the PS
empirical scaling approach and the green dash-dotted line from Ziegler's SRIM
program.}}
\label{fig:stoppingPowerComparison}
\end{figure}

Fig.~\ref{fig:stoppingPowerComparison} gives the electronic stopping power
for Ar ions in Ar targets. The red solid line is produced based on BK model
and the ``heavy ion scaling rule''. Paul and Schinner (PS)~\cite{Pau01} 
performed a 
systematic comparison of a large collection of stopping power data for
projectiles from $^3$Li to $^{18}$Ar to those for alpha particles in the
same materials.  They found that, for each target element, a slightly
adapted sigmoid function with three parameters can describe 
the normalized relative stopping power
\begin{equation}
S_{\rm{rel}} = \frac{S_A/Z_1^2}{S_{\rm {He}}/2^2}. 
\label{eq:S_rel}
\end{equation}  

\noindent
For comparison, results from the PS empirical scaling approach 
have also been plotted in Fig.~\ref{fig:stoppingPowerComparison} 
with a blue dashed line. The difference between the two calculations 
is about 10\% at kinetic energy
of 1 MeV and increases to 35\% at 10 keV. In addition, a green 
dash-dotted line in the figure shows a calculation based on the ``SRIM'' 
program of Ziegler~\cite{zbl}. The results deviate from those of the
BK and PS approaches at higher energy with a kink at $\sim 0.1$ MeV, but are 
comparable to the results of the BK model at lower energy.

\begin{figure}[htb!!!]
\includegraphics[angle=0,width=12.cm] {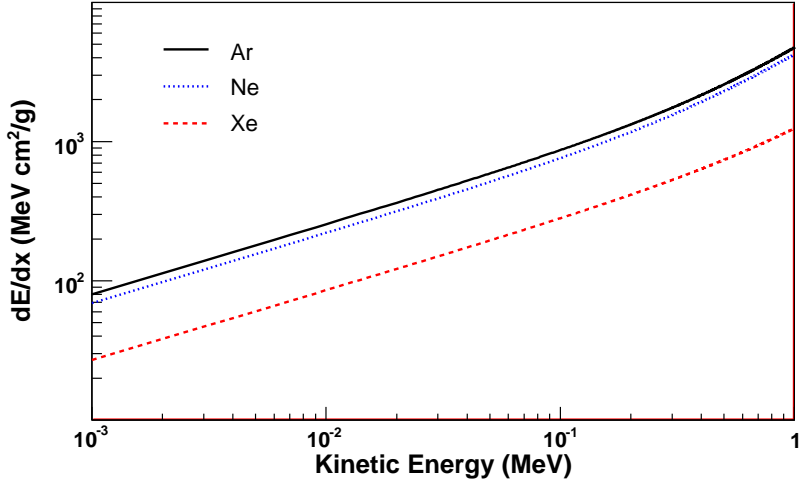}
\caption{\small{
The electronic stopping powers based on the BK model for Ar (black solid line),
Ne (blue dotted line) and Xe (red dashed line)
ions with themselves as the target materials.}}
\label{fig:stoppingPower_Ar_Ne_Xe}
\end{figure}

Based on the BK model, we show in Fig.~\ref{fig:stoppingPower_Ar_Ne_Xe} 
the electronic stopping power for Ar (black solid line), 
Ne (blue dotted line) and Xe (red dashed line) 
ions with themselves as the target materials. By comparing 
different effective ion charge calculation scenarios, 
the theoretical uncertainties
on the stopping power values are estimated to be about 15\% at 1 MeV 
and increase to about 50\% at 1 keV where the theoretical description 
is known to become less reliable. 

\section{Model Prediction}
\label{sec:com}

The fraction of nuclear recoil energy, $f_{n}$  
(see Eq.~\ref{eq:quen}), described in Section \ref{sec:quen1} 
ionizes or excites atoms in noble liquids. Ionizing particles produce 
excitons, electron-ion pairs,  and a localized concentration core along the track. There are two cases: 1) free excitons form 
excimers through a self-trapping process~\cite{ahi, atd, tta}, 
and 2) 
the ions are localized 
through the formation of excited molecular ions and eventually form excimers through
recombination, deexcitation, and collision. 
The origins of the
luminescence for both cases are attributed~\cite{rsm,mma} to low excited molecular states, namely,
$^{1}\Sigma_{u}^{+}$ or $^{3}\Sigma_{u}^{+}$. 
The fraction of excitons that undergo self-trapping depends largely on ionization density. The fraction
of ions that combine with ground states to form excited molecular ions is proportional to ionization
density. Furthermore, the fraction of excimers relaxing to the lowest-excited states is also a function of
ionization density. This is to say that some of the excitons or ions can collide each other without 
going to excimers and some of the formed excimers can further undergo the Penning process
to reduce the number of radiative lowest-excited states. This part of excitons, ions, and the formed excimers is quenched. 
The quenching factor, $f_{l}$,  
is  
expressed in  Eq.~\ref{eq:birk2}.  Since $f_{n}$ and 
$f_{l}$ are independent of each other, the total scintillation efficiency in noble liquids 
can be represented by 
 
\begin{equation}
\label{eq:quenf}
q_f = f_{n} \times f_{l}.
\end{equation} 

We calculated $f_{n}$, the fraction of nuclear recoil energy that contributes 
to the ionization process in noble liquids, as a function of nuclear 
recoil energy, as shown in 
Fig.~\ref{fig:lindfirst}. 
The fraction of the collided excitons or ions, $kB$, 
is determined from experimental data based on Birk's 
saturation law by using Eq.~\ref{eq:birk2}.

\begin{figure}[htb!!!]
\includegraphics[angle=0,width=12.cm] {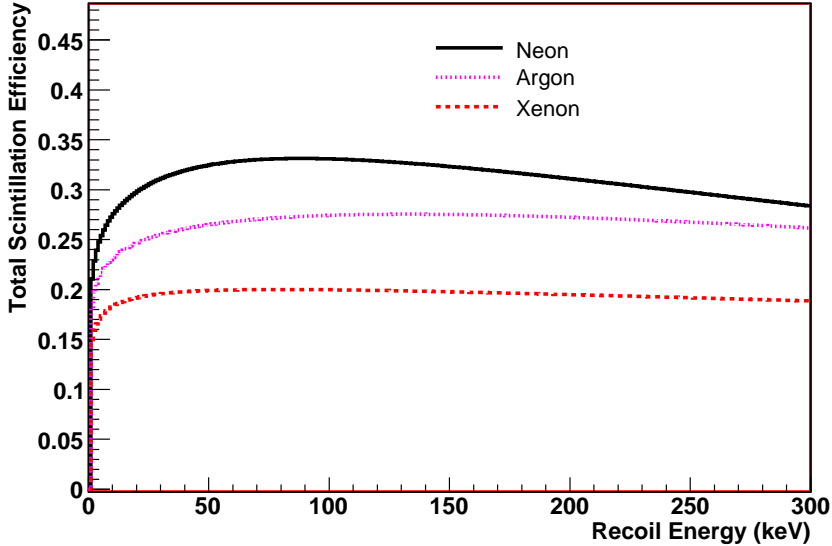}
\caption{\small{
The total scintillation efficiency for nuclear recoils as a function of recoil
energy in liquid neon (black line), argon (red dotted line), and xenon (magenta dashed line).}}
\label{fig:argon1}
\end{figure}

\begin{itemize}
\item{For liquid argon,  $kB = 7.4 \times 10^{-4}$ MeV$^{-1}$ g cm$^{-2}$. 
This is determined by using a  quenching factor (46\%) from a heavy ion 
measurement~\cite{ion}, and $dE/dx$ (1586.4 MeV cm$^{2}$/g 
corresponds to 31.9 MeV/amu) is calculated by this work.}
\item{For liquid neon, the Birk's constant, 
$kB= 1.12 \times 10^{-3}$ MeV$^{-1}$ g cm$^{-2}$, is determined by
using a quenching factor measured in Ref.~\cite{ne} with 
the $dE/dx$ values calculated by this work. }
\item{For liquid xenon, the quenching factor has been 
measured by several groups~\cite{xe1,xe2,xe3}. We use the 
measured quenching factor corresponding to a 70-keV recoil energy 
in Ref.~\cite{xe2} to determine the Birk's constant $kB$ to 
be 2.015 $\times 10^{-3}$ MeV$^{-1}$ g cm$^{-2}$. }
\end{itemize}
The quenching factor due to scintillation quenching, $f_{l}$, is then calculated 
according to Eq.~\ref{eq:birk2} for argon, neon and xenon.
Multiplying these two reduction factors 
results in the total scintillation efficiency (Eq.~\ref{eq:quenf}) as shown in 
Fig. \ref{fig:argon1}.

\section{Comparison of model prediction with data}
\label{sec:comparison}

\begin{figure}[htb!!!]
\includegraphics[angle=0,width=12.cm] {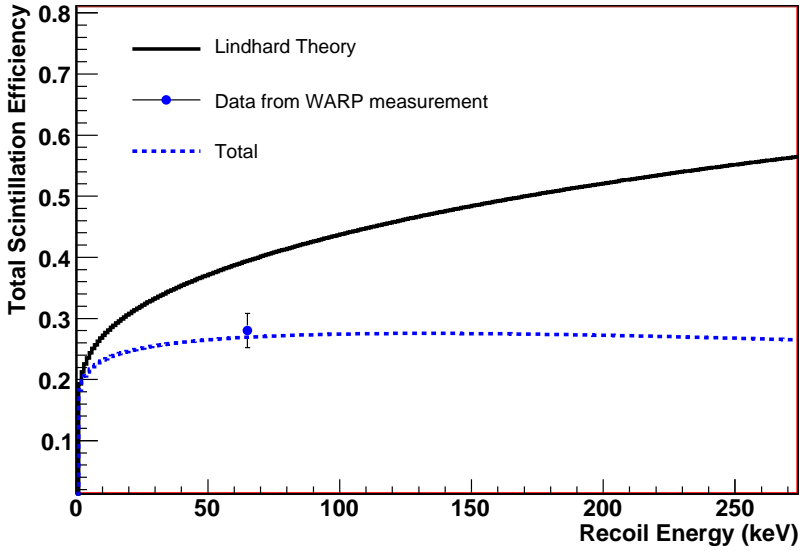}
\caption{\small{
Argon: a comparison of the total scintillation efficiency (blue curve) calculated 
by combining two reduction factors to the experimental data~\cite{ar} 
in liquid argon. Also shown as the black curve is the Lindhard ionization energy reduction factor.}}
\label{fig:argon2}
\end{figure}

\begin{figure}[htb!!!]
\includegraphics[angle=0,width=12.cm] {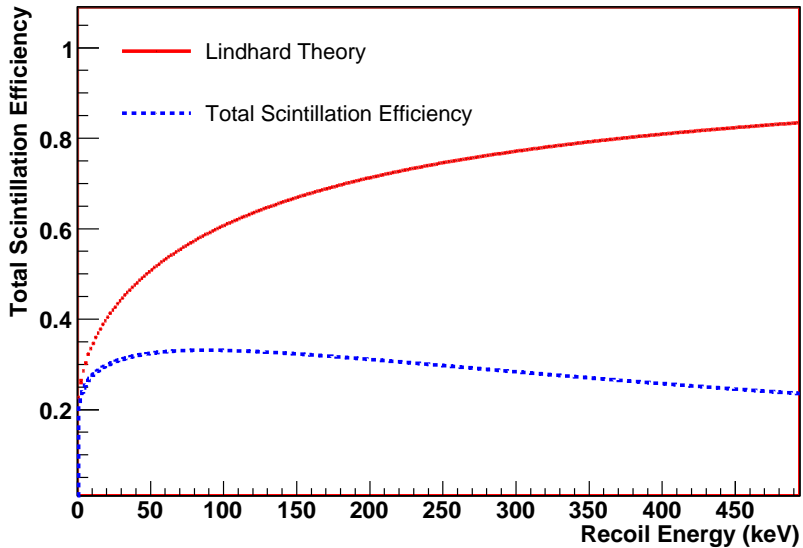}
\caption{\small{
Neon: the total scintillation efficiency (blue curve) calculated
by taking into account two reduction factors is presented 
in liquid neon. }}
\label{fig:neon2}
\end{figure}

\begin{figure}[htb!!!]
\includegraphics[angle=0,width=12.cm] {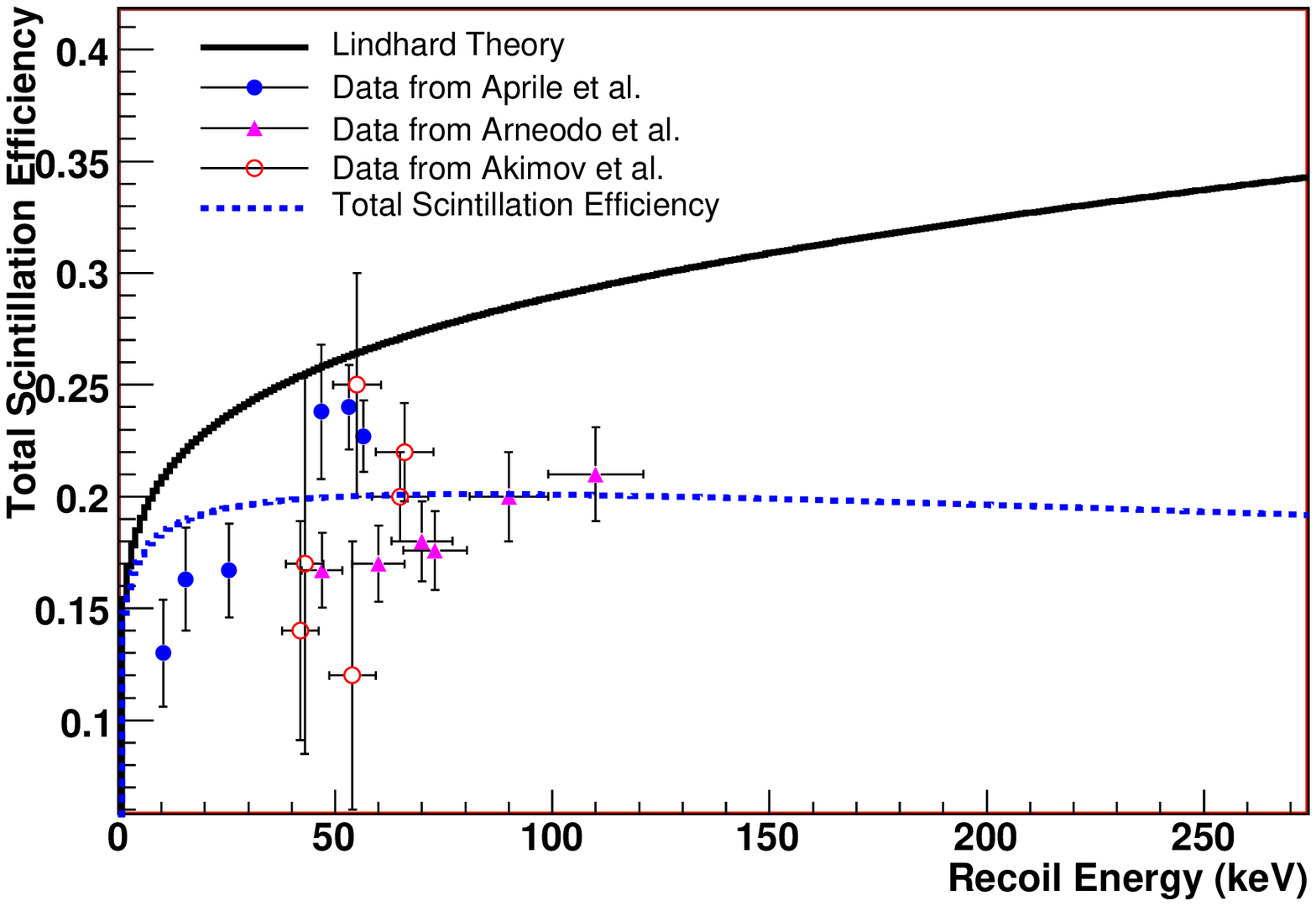}
\caption{\small{
Xenon: a comparison of the total scintillation efficiency (blue curve) calculated
by combining two reduction factors to the experimental 
data~\cite{xe1, xe2, xe3}
in liquid xenon. Also shown as the black curve is the Lindhard reduction factor.
}}
\label{fig:xenon2}
\end{figure}

The total scintillation efficiency for each noble liquid, argon, neon, and xenon, is  
 compared to experimental 
data in Fig. \ref{fig:argon2},  Fig.~\ref{fig:neon2}, and  Fig. \ref{fig:xenon2}. 
As can be seen, the experimental data can not
be explained by Lindhard's theory alone, but can be well described 
when the Birk's saturation effect has been taken into account. Note that the scintillation efficiencies for 5.035 MeV $\alpha$ 
particles and 33.5 MeV/amu $^{18}$O ions in liquid argon were measured to be 71\% and 59\%~\cite{nim, ion}, which agree very well with our 
model prediction, 72\% and 63\%.

It is worthwhile to mention that, although the theoretical 
uncertainties of electronic stopping power values at very low 
recoil energy are considerable, the effect on the uncertainties
of the final quenching factor are not large as the Birk's constants
$kB$ are quite small. For example, by assuming a factor of 
two difference in the electronic stopping powers at 1 keV in liquid argon, 
the contribution to the uncertainty of the quenching factor is less than 7\%. 
Thus, the dominant uncertainties come from the uncertainties
in the Lindhard theory at very low energy. However, it is usually 
believed that the Lindhard theory describes experimental data well~\cite{ars,cch, tsc, gge}, 
although  the experimental uncertainties are quite large at 
low energy.  

\section{Low-Energy Response to Electrons and $\gamma$-rays}
\label{sec:ele}
\begin{figure}[htb!!!]
\includegraphics[angle=0,width=12.cm] {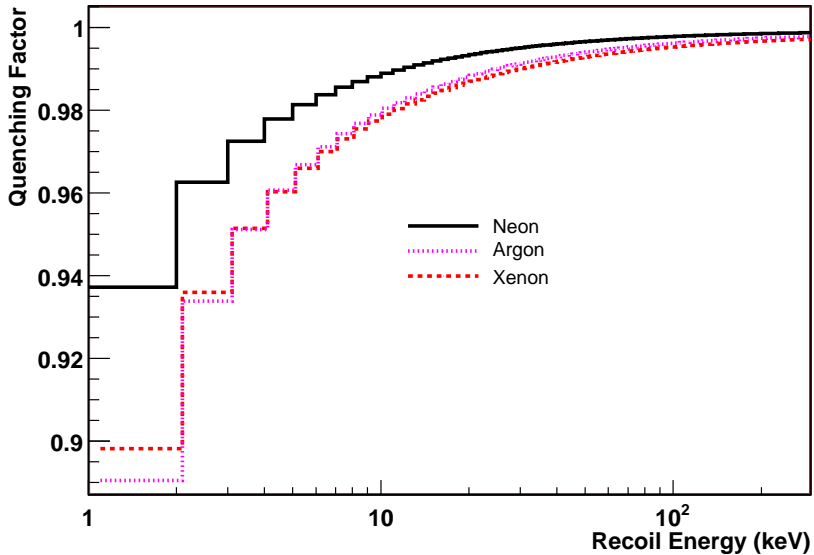}

\caption{\small{
The quenching factor calculated for electronic recoils in
liquid argon (red dotted line), neon (black line), and xenon (magenta dashed line).
}}
\label{fig:ele1}
\end{figure}

Noble liquids' response to low-energy electrons and $\gamma$-rays is 
usually assumed to be linear for the existing 
experiments~\cite{xenon10, warp, dan}. To investigate whether 
such an assumption holds, we also calculate and show in Fig.~\ref{fig:ele1} the quenching factor
for electronic recoils in liquid
argon, neon and xenon utilizing the $kB$ constants determined above. It is clear that the assumption 
holds quite well down to 20 keV because of the small stopping power 
values (a few MeV cm$^{2}$/g) above 20 keV. However, the quenching effect is 
expected to be significant when the electronic
recoil energy is below 20 keV due to the larger stopping power (more than 10 MeV cm$^{2}$/g). 

Note that the $kB$ constants
are determined by Birk's law in which the relative scintillation efficiency of nuclear recoils is used, under the assumption that 
the electronic recoil quenching factor is 1.  As we can see in Fig.~\ref{fig:ele1}, this assumption is only valid above 20 keV.      

\section{Conclusion}
In summary, we found that the reduced scintillation response to low-energy 
nuclear recoils comes from two different mechanisms: 
1) reduced ionization and excitation energy by nuclear collisions and 
2) reduced scintillation photon yield due to high ionization and excitation density 
induced by nuclear recoils. The former is well described by Lindhard's theory and the latter  
is attributed to biexcitonic collisions between excitons and the Penning process between excimers. The scintillation quenching
induced by the combination of biexcitonic collisions and the Penning process  can be described by Birk's law.
 We have combined these two reduction mechanisms 
to describe the total scintillation efficiency for noble liquids. The calculations are 
compared to available data and it is found that they are in good agreement within
experimental uncertainties. We have also studied the scintillation response of
noble liquid scintillators to low-energy electrons and $\gamma$-rays and
found that the response is not linear at recoil energy below 20 keV. This paper
provides a conventional way to measure the total scintillation efficiency of noble liquids for all types of particles
including neutron-induced nuclear recoils, alpha particles and other heavier elements
by measuring Birk's constant $kB$ using $\gamma$-ray sources.
\label{sec:con}

\section*{Acknowledgments}
The authors wish to thanks to the research group at University of
South Dakota for the invaluable support that made this work successful. 
This work was supported in part by the Office of Research at University 
of South Dakota and by Laboratory Directed Research 
and Development at Los Alamos National Laboratory. Z.Y. 
was also partly supported by MOE of China under project No. IRT0624 
and the NSFC under grant No. 10635020.

%
%

\end{document}